\documentclass[twocolumn]{aastex61}
\usepackage[english]{babel}
\usepackage{graphicx}
\usepackage{amsmath,amssymb,latexsym}
\usepackage{natbib}
\usepackage{bm}
\usepackage{calrsfs}
\usepackage{hyperref}
\usepackage{ulem}
\usepackage{xspace}
\begin{document}
\def\be{\begin{equation}}
\def\ee{\end{equation}}
\def\ba{\begin{eqnarray}}
\def\ea{\end{eqnarray}}
\def\d{\delta}
\def\e{\epsilon}
\def\f{\varphi}
\def\k{\varkappa}
\def\tde{\tilde}
\def\p{\partial}
\def\ms{\mathstrut}
\def\s{\strut}
\def\ds{\displaystyle}
\def\ts{\textstyle}
\def\b{\boldsymbol}
\def\r{\mathrm}
\def\G{\Gamma}
\def\sun{\odot}
\def\ergs{\ensuremath{\mathrm{erg\,s^{-1}}}}
\def\erg{\ensuremath{\mathrm{erg}}}
\def\ag{{\it AGILE}\xspace}
\def\fer{{\it Fermi} LAT\xspace}
\def\magic{MAGIC\xspace}
\def\hess{H.E.S.S.\xspace}
\def\veritas{{\it Veritas}\xspace}
\def\cta{{CTA}\xspace}
\def\nustar{{\it NuSTAR}\xspace}
\def\pwn{\protect\object{HESS J1825$-$137}\xspace}
\def\psr{\protect\object{PSR~J1826$-$1334}\xspace}
\def\dst{\displaystyle}
\def\cos{\mathop{\rm cos}\nolimits}
\def\sin{\mathop{\rm sin}\nolimits}
\def\tg{\mathop{\rm tg}\nolimits}
\def\ctg{\mathop{\rm ctg}\nolimits}
\def\ln{{\rm ln}}
\def\const{{\rm const}}
\def\div{{\rm div}}
\def\min{{\rm min}}
\def\max{{\rm max}}
\def\dfrac#1#2{{\displaystyle#1\over\displaystyle#2}}
\def\prodi#1#2{\frac{d #1}{d #2}}
\def\prodii#1#2{\frac{d^2 #1}{d {#2}^2}}
\def\parti#1#2{\frac{\partial #1}{\partial #2}}
\title{On the anomalously large extension of the Pulsar Wind Nebula  \pwn}
\correspondingauthor{Dmitry Khangulyan}
\email{d.khangulyan@rikkyo.ac.jp}
\def\dias{\affiliation{Dublin Institute for Advanced Studies, School of Cosmic Physics, 31 Fitzwilliam Place, Dublin 2, Ireland}}
\def\mephi{\affiliation{National Research Nuclear University (MEPHI), Kashirskoje shosse, 31, Moscow, Russia}}
\def\mpik{\affiliation{Max-Planck-Institut f\"ur Kernphysik, Saupfercheckweg 1, D-69117 Heidelberg, Germany}}
\def\rikkyo{\affiliation{Department of Physics, Rikkyo University, Nishi-Ikebukuro 3-34-1, Toshima-ku, Tokyo 171-8501, Japan}}
\def\riken{\affiliation{RIKEN iTHEMS, Hirosawa 2-1, Wako, Saitama 351-0198, Japan}}
\def\mphti{\affiliation{Moscow Institute of Physics and Technology, Institutskiy per. 9, Dolgoprudny, Russia}}
\def\ipm{\affiliation{Keldysh Institute of Applied Mathematics RAN, Miusskaya sq. 4, Moscow, Russia}}
\author[0000-0002-7576-7869]{Dmitry Khangulyan}
\rikkyo
\riken
\author{Alexander V. Koldoba}
\mphti
\author{Galina V. Ustyugova}
\mphti
\ipm
\author{Sergey V. Bogovalov}
\mephi
\author{Felix Aharonian}
\dias
\mpik
\mephi
\begin{abstract}
The very high energy (VHE) gamma-ray emission reported from a number of pulsar wind nebulae (PWNe) is naturally
explained by the inverse Compton scattering of multi-TeV electrons. However, the physical dimensions of some gamma-ray-emitting PWNe
significantly exceed the scales anticipated by the standard hydrodynamical paradigm of PWN formation.  The most
``disturbing'' case in this regard is HESS~J1825-137, which extends to distances
$r\approx70\rm\,pc$ from the central pulsar \psr.  If the gamma-ray emission is indeed produced inside the PWN, but not
 by electrons that escaped the nebula and diffuse in the interstellar medium (ISM), the formation
of such an anomalously extended plerion could be realized, in a diluted environment with the hydrogen
number density $n_\textsc{ism}\le10^{-2}\rm\,cm^{-3}$.  In this paper, we explore an alternative scenario
assuming that the pulsar responsible for the formation of the nebula initially had a very short rotation period. In
this case, the sizes of both the PWN and the surrounding supernova remnant depend on the initial pulsar period, the
braking index, and the ISM density. To check the feasibility of this scenario, we study the parameter space
that would reproduce the size of HESS~J1825-137.  We show that this demand can be achieved if the braking index is
small, $n\leq2$ and the pulsar birth period is short, $P_{\rm b}\simeq1\rm\,ms$. This scenario can
reproduce the wind termination position, which is expected at \(R_\textsc{ts}\simeq0.03\rm\,pc\), only in a dense environment with $n_\textsc{ism}\geq\rm1\,cm^{-3}$.  The requirement
of the dense surrounding gas is supported by the presence of molecular clouds found in the source vicinity.
\end{abstract}
\section{Introduction}
Pulsar wind nebulae (PWNe) constitute a large nonthermal source population, many representatives of which are prominent
emitters of gamma-rays, especially in the very high energy (VHE; $E_\gamma \geq 100$~GeV) band \citep[see, e.g.,][and
references therein]{2015SSRv..191..391K}.  These objects are formed in the course of the interaction of the pulsar wind
with the surrounding matter -- the interstellar medium (ISM) or the interior of the related supernova remnant (SNR). At
this interaction, a substantial fraction of the pulsar's spin-down energy is transferred to ultrarelativistic electrons
with energies extending to \(100 \,\rm TeV\) and beyond.  The  subsequent interactions of these electrons with the ambient photon (basically, the microwave background radiation, MBR) and magnetic fields  result in  the  
formation of synchrotron and inverse Compton
(IC) nebulae in the X-ray and high-energy gamma-ray bands.  
The energy losses of these electrons are shared between the synchrotron 
and IC radiation channels in a proportion determined by the energy densities of the magnetic field, \(w_\textsc{b}\), and the photon target, \(w_\textsc{mbr}\):  $L_\gamma / L_{\rm X} = w_\textsc{mbr}/w_\textsc{b} \approx 1 (B/ 3 \mu \rm G)^{-2}$.  Thus,  for most of the PWNe 
with a magnetic field of the order of $10 \ \rm \mu G$ or less the gamma-ray production efficiency could exceed 10 \%.  Based on the assumption that a major fraction of the 
rotational energy of pulsars  is released in ultrarelativistic 
electrons accelerated at the wind termination shock (TS), 
it has been predicted \citep{1997MNRAS.291..162A} that the ground-based detectors with the performance typical for the current Atmospheric Cherenkov Telescopes should be able to 
reveal VHE gamma-ray component of radiation  of tens of PWNe with   spin-down luminosities  $L \geq 10^{34} (d/1\rm\, kpc)^2\,\rm \ergs$ ($d$ is the distance to the source).  The detection of almost three dozen of extended VHE gamma-ray sources in the vicinity of pulsars \citep[see, e.g.,][and the TeVCAT catalog\footnote{http://tevcat.uchicago.edu/}]{2018A&A...612A...1H,2018A&A...612A...2H} 
in the indicated range of spin-down luminosities supports this prediction. 
Thanks to the relatively 
large field of view,  the good energy and angular resolutions, and the 
vast collection areas, the current atmospheric Cherenkov gamma-ray
telescopes allow deep studies of energy-dependent morphologies of PWNe  on angular 
scales of  $\sim 1^\circ$.  Sources with larger angular extension can be studied with HAWC, although with lower spatial and energy resolutions \citep{2017ApJ...843...40A,2017PhRvD..96j3016L}.  The potential of
X-ray instruments in this regard is relatively modest; see, however, \citet{2010ApJ...719L.116B} for a study of extended PWNe in the X-ray band.

Gamma-rays being the product of IC
scattering of electrons provide unambiguous model-independent information about the energy and spatial distribution of
parent electrons. 
While the spectral energy distributions (SEDs) of X-rays and gamma-rays from most of the PWNe are comfortably
described within the current models of PWNe, the reported TeV gamma-ray images extending to $\sim 1^\circ$ do not have a
simple explanation.  In particular, the angular size of \pwn, a PWN associated with the pulsar \psr, located at a distance
of $\sim 4$~kpc, corresponds to the physical extension of the gamma-ray production region  of \(70\rm\,pc\)
\citep{2006A&A...460..365A,2017AIPC.1792d0035M,2017arXiv170803126M,2018A&A...612A...2H}.
Indeed, the particle acceleration in PWNe is presumably linked to the pulsar wind termination, which in \pwn is expected to occur at a relatively small distance from the pulsar, \(R_{\textsc{ts}}<0.1\rm\,pc\)  \citep[see, e.g.,][]{2011ApJ...742...62V}.  The formation of such an extended TeV source requires either operation of a highly efficient particle transport or {\it in situ} particle (re)acceleration. There are two primary transport mechanisms for high-energy particles in PWNe: (i) advection by the macroscopic flow and (ii) diffusion.
The characteristic propagation times  depend strongly  on the properties  of the background plasma, which  is determined, to a large extent, by the hydrodynamic flow. Thus, a consistent hydrodynamical description of PWNe is an essential element for understanding of the transport of relativistic particles in these objects.
In this paper, we explore  the formation of extended PWNe powered by a powerful pulsar in the context of hydrodynamics (HD) of the  PWN--SNR system. The size of PWN determines the region where relativistic particles can advect with the nebular flow.  As the zeroth-order approximation, we ignore the dynamical impact of the magnetic field.  The magnetic field in extended PWNe is typically small, but still, the magnetic pressure can be comparable to the particle pressure \citep[see][for the case of \pwn]{2011ApJ...742...62V}.  However, if one focuses on the expansion of a PWN, the main effect comes from the equation of state, which is typically taken as a polytropic with the index \(4/3\) for both the magnetized and nonmagnetized relativistic plasma.  The magnetic field can determine the preferred direction for the PWN expansion,  as seen, for example, in the Crab Nebula. However, in the case of very extended PWNe, the nonhomogeneity of the ISM renders a much stronger influence on the PWN shape.  The pure HD model can provide a simple, but still  quite correct  description for extended PWNe, without invoking the impact of the magnetic field.
\begin{figure}
\resizebox{\columnwidth}{!}{
\includegraphics[]{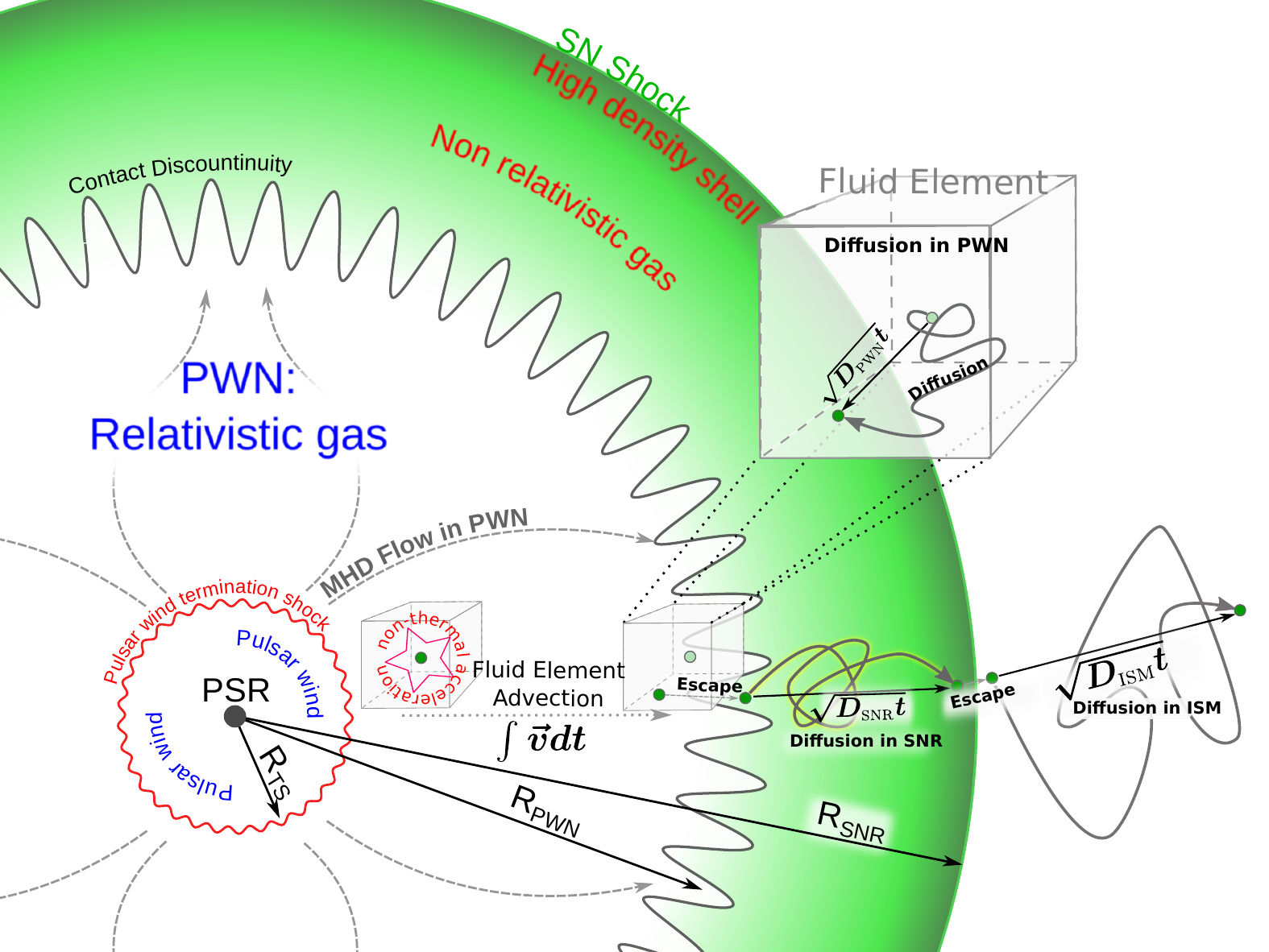}}
\caption{Schematic representation of the main zones and the 
particle transport mechanisms in composite SNRs. }
\label{fig:sketch}
\end{figure}
\section{A Model for PWN in the Composite SNR}\label{sec:pwn}
SNRs that contain a PWN -- the so-called composite SNRs -- have a specific structure, which is presented schematically in
Fig.~\ref{fig:sketch}.  The part that is referred to as PWN is filled with relativistic particles that originate from the
pulsar. Bulk velocities in this region can be relativistic, in particular, close to its inner boundary, which is the
pulsar wind TS. The contact discontinuity (CD) bounds the PWN from outside. The plasma velocities close
to the CD are nonrelativistic. 
Thus, an efficient advection in PWNe is most
likely  limited by  the inner region of the nebula, which may, however, extend significantly beyond the TS.
In contrast to advection, the diffusive transport is not constrained by any specific hydrodynamic region. It
allows particles to escape from the PWN to the SNR region. Apparently, the diffusion coefficients in PWN and SNR may differ significantly.
Similarly, the advection in the SNR is limited by the region between the CD and the SN shock. In this region, the flow
velocities are small (\(\sim10^4\rm\,km\,s^{-1}\)).  The diffusion in the SNR leads to the particle escape.  The high-energy particles from the SNR can penetrate PWN or eventually escape to the ISM.  In the ISM, the diffusion proceeds at a
rate corresponding to the diffusion of the galactic cosmic rays (CRs).  Note that in the Galactic disk the CR diffusion
is much (orders of magnitude) faster compared to the diffusion in the Bohm limit.  Such a diffusion might be a key element for the formation of extended TeV sources around pulsars \citep{1995NuPhS..39..193A,2017Sci...358..911A}. Finally, advection in ISM might be
relevant only in the case of high proper velocity of the pulsar.

The evolution and the structure of PWNe are essential for understanding of the origin of radiation seen from these objects.  The
study of various aspect of PWNe has a long history. \cite{1973ApL....13..103P} studied the evolution of the magnetic
field at the early epoch of formation of the Crab Nebula and explored the impact of this evolution on the distribution
of nonthermal particles and strength of the magnetic field at the present epoch \citep{1973ApJ...186..249P}.  The
impact of the dynamics of PWNe on a self-gravitating shell was investigated by \citet{1974Ap&SS..28..205M}.  Later on,
\citet{1984ApJ...278..630R} proposed a comprehensive model describing the dynamics of evolution of the PWNe in the SNR
\citep[see also][]{2011MNRAS.410..381B}.
In some other studies, the structure of PWNe within a slowly expanding SNR shell has been explored
\citep{1974MNRAS.167....1R}. This approach has been developed into a detailed 1D magnetohydrodynamic (MHD) model for the
Crab Nebula \citep{1984ApJ...283..694K,1984ApJ...283..710K}.  The latter describes remarkably  well the broadband SED of the
Crab Nebula from the optical to soft gamma-ray band through the synchrotron radiation and in the VHE band through the IC
scattering of electrons \citep{1996MNRAS.278..525A}.
Finally, in recent years different aspects of the PWN evolution have been addressed through 2D and 3D (M)HD numerical
simulations. In particular, a significant progress has been achieved in the modeling of X-ray morphology caused by the
pulsar wind anisotropy \citep[see, e.g.,][]{2004A&A...421.1063D,2014MNRAS.438..278P} and by the reverse shock at the
later stages of evolution of the PWN \citep[see, e.g.,][]{2017ApJ...844....1K}.

In this paper, we develop an analytical model to describe the dynamics of evolution of the anomalously extended PWNe.
Generally, the studies of PWNe silently assume that the dominant contribution to the energy of a composite SNR is
provided by the supernovae (SN) explosion.  Indeed, the total deposited energy by one of the currently most powerful
pulsars, the Crab pulsar, is estimated as \(3\times10^{50}\erg\), which is significantly smaller than the energy
transferred to the ejecta at the SN Type II explosion \citep{1984ApJ...278..630R}.  If the SNR energy is dominated by
the SN explosion, then Sedov's solution provides a rather accurate estimate of the source size.  In this
case, extended PWNe, with a size significantly exceeding \(10\rm\,pc\), are possible only in regions
characterized by an extremely low density of the ISM \citep[e.g.,][]{2009ASSL..357..451D}.  The required density, 
as low as $n_\textsc{ism} \approx 10^{-3} \ \rm cm^{-3}$,  is not typical for the 
conventional  regions of the ISM in the Galactic Plane. Moreover, 
in the specific case of \pwn, it is  not supported by observations that revealed dense 
molecular clouds  in the vicinity of  \pwn \citep{2016MNRAS.458.2813V}.
Below, we explore whether the problem can be resolved (or at least relaxed) in the case of injection of a large amount of energy by the pulsar
that dominates over the SN explosion energy.  In this case, the dynamics of the system should be similar to {\it
  Phase IV} of the scheme discussed by \citet{1984ApJ...278..630R}.  However, it is important to explicitly account for
the dynamics of the ISM involved in the motion.  Indeed, the mass of interstellar matter located in a
region with a radius \(R_{\textsc{snr}}=10\rm\,pc\) can achieve \(100M_\sun\) for the mean proton density of
\(1\rm\,cm^{-3}\).  We propose a dynamic model that accounts for these two important factors, the dominant pulsar
contribution to the dynamics of SNR, and the inertia of the ISM.
\subsection{The contribution of pulsar to the overall SN energy}
The pulsar rotation losses are determined through the time dependence of its angular velocity: 
\(
\dot{E}=I\Omega\dot{\Omega}\,,
\)
where \(\Omega\), \(\dot\Omega\), and \(I\) are the pulsar angular velocity, its time derivative, and the moment of inertia of the neutron star (NS), respectively. The later is calculated using the relation
\be
I\simeq10^{45}{\,\rm\,g\, cm^2}\,\left[\left({M_{\textsc{psr}}\over 1.4 M_\odot}\right)\left({R_{\textsc{psr}}\over 10 \rm\,km}\right)^{2}\right]\,.
\ee 
It is conventionally assumed that the change of the angular velocity is determined by the braking index, \(n\):
\(
\dot{\Omega}\propto\Omega^n\,.
\)
If the  time  \(t'=0\) corresponds to the present epoch, the pulsar power evolves with time as
\be
L_{\rm sd}(t)={L_{p}\over \left(1+t'/t_0'\right)^{{n+1\over n-1}}}\,.
\ee
Here, \(\Omega_{p}\) and \(\dot{\Omega}_{p}\) are the current angular velocity and the angular acceleration, respectively. The {\itshape characteristic} time, \(t_0'\) is determined by the pulsar {\it characteristic}  age, \(\tau=-\Omega_{p}/2\dot\Omega_{p}\), and the braking index,
\be
t_0'={2\tau\over (n-1)}\,.
\ee
In the above treatment,  the pulsar birth epoch, \(-t_0'<t_{\rm b}'<0\),  is  uncertain.  For almost all pulsars, the measured values of the braking index, \(n\),  vary  between \(1\) and \(3\) \citep{1993MNRAS.265.1003L,1995ApJ...448..365B,1996Natur.381..497L,2007Ap&SS.308..317L,2011MNRAS.411.1917W,2011ApJ...730...66L,2012MNRAS.424.2213R}, with  two exceptions: PSR~J1734\(-\)3333 
 \citep[\(n=0.9\pm0.2\),][]{2011ApJ...741L..13E}  and PSR~J1640\(-\)4631 \citep[\(n=3.15\pm0.03\),][]{2016ApJ...819L..16A}. Thus, the true age of the pulsar can 
exceed,  by a  factor  \(>5\), the characteristic age, \(\tau\) \citep[see, e.g.,][]{2005ICRC....4..239D}.
The birth angular velocity is constrained by the centrifugal breakup limit,
\be
\begin{split}
\Omega<\Omega_{\rm br}&=\sqrt{G M_{\textsc{psr}}\over R_{\textsc{psr}}^3}\\
&=\Omega^{(0)}_{\rm br}\left[\left({M_{\textsc{psr}}\over 1.4 M_\odot}\right)^{1/2}\left({R_{\textsc{psr}}\over 10 \rm\,km}\right)^{-3/2}\right]\,,\\
\end{split}
\ee
where \(\Omega^{(0)}_{\rm br}=1.4\times10^4 {\rm\,s^{-1}}\) corresponds to the breakup velocity for the conventional size of the NS.

The entire energy released by the pulsar since its birth at \(t_b\) is 
\be
E=\int\limits_{t_b'}^0 dt' L_{\rm sd}(t')={I\left(\Omega_{\rm b}^2-\Omega_{\rm p}^2\right)\over 2}\,,%L_{p}\tau\left(\left({\Omega_{\rm b}\over \Omega_{\rm p}
\ee
where \(\Omega_{b}\) is the birth angular velocity.
Unless, the mass and the size of the NS strongly deviate from their conventional values, 
the breakup constraint limits  the total energy released by the pulsar: \(E<10^{53}\erg\).
Note that $E$  does not depend  on the pulsar braking index \(n\). This estimate ignores, however, the effects related to the dependence of the moment of inertial on pulsar's angular velocity. For example, for the velocities close to the breakup limit, the shape of the NS should deviate significantly from the spherical one, resulting in the dependence of \(I\) on \(\Omega\) \citep[see, e.g.,][]{2015PhRvD..91f3007H}.  The evolution of the NS magnetic field and direction of the magnetic momentum can also lead to a nontrivial dependence of the braking index on time \citep[see, e.g.,][]{2016PhRvD..94f3012H,2016MNRAS.457.1180R,2017MNRAS.465..383R}.  The consideration of these effects is beyond the scope of this paper.

{ Although  a rigid rotating sphere does not, strictly speaking, represent an accurate model for an NS, when its angular velocity is close to the breakup limit, for the sake of simplicity we still use this approximation. In this case, the rotation energy of the pulsar is}
\be\label{eq:e_rot}
\begin{split}
  E_\textsc{rot}&={I\Omega_{\rm br}^2\over 2}=\frac15{GM^2\over R}\\
  &\simeq10^{53}\ergs\left[\left({M_{\textsc{psr}}\over 1.4 M_\odot}\right)^{2}\left({R_{\textsc{psr}}\over 10 \rm\,km}\right)^{-1}\right]\,.
\end{split}
\ee
{ The rotation energy constitutes a substantial fraction of the gravitational energy released during the formation of the compact object:} \(E_\textsc{grav}=3E_\textsc{rot}\) {  (here, both estimates ignore the relativistic effects). The formation of a pulsar at SN explosion  \citep[see, e.g.,][]{1999ApJ...513..780P} does not  a priory exclude that the pulsar gets an angular velocity close to the breakup limit \citep[see, e.g.,][]{2005ApJ...626..350H}.   Moreover, the bulk of the gravitational energy released during the formation of the compact object is expected to escape with the neutrino emission, and only a tiny fraction, \(\sim1\%\), is transferred to the ejecta,} \(E_{\textsc{ej}}\sim10^{51}\erg\). { Thus,  even if the pulsar rotation energy at its birth is small compared to the breaking limit, it  still can exceed the initial kinetic energy of the SN shell. In this case, the dimensions of the SNR and the PWN are  determined by the pulsar injection energy. }

\subsection{Structure of  the Composite SNR}
The expansion of the PWN determines the evolution of  distribution of 
nonthermal particles  and the strength of the magnetic field. 
Given the relativistic nature of plasma in PWNe, at the stages relevant for the modeling of the present-day nonthermal emission, the hydrodynamic processes in the PWN should proceed in the nearly adiabatic regime. Thus, the internal structure of the nebula can be described analytically. On the other hand, the state of the nonrelativistic shell may depend strongly on its dynamics at earlier stages of the expansion. To explore this evolution, we use a simple dynamic model.
The energy of the system is distributed between three components: the kinetic energy of the shell (SNR shell), the internal energy of the nonrelativistic gas in the shell, and the internal energy of the relativistic gas in the nebula. The SNR shell occupies the region between the external boundary of the PWN  at \(R_{\textsc{pwn}}\), and the SNR radius, \(R_{\textsc{snr}}\).  The mass of the SNR shell is determined by the gas density of the ISM, \(\rho_{\textsc{ism}}\), and the shell radius. The SNR density varies strongly throughout the shell: the mass is concentrated in a thin layer at \(R_{\textsc{snr}}\), while the hot nonrelativistic gas fills the remaining part of the shell. The pressure in the hot part of the shell, \(P\), is equal to the pressure of the relativistic gas in the PWN.
To describe the system, we use a model that includes the following relations:
\begin{enumerate}
\item the equation for the energy balance in the nebula,
\be \label{eq:eq1}
3{d\,PV_{\textsc{pwn}}\over d\,t}+P{d\,V_{\textsc{pwn}}\over dt}=L_{\rm sd}(t)\,;
\ee
\item the equation for the energy balance in the nonrelativistic shell,
\be \label{eq:eq2}
{d\over d\,t}\left({Mv^2\over2}+\frac32{P(V_{\textsc{snr}}-V_{\textsc{pwn}})}\right)=P{d\,V_{\textsc{pwn}}\over d\,t}\,,
\ee
where $v$ is the shell velocity; and
\item the equation of motion,
\be \label{eq:eq3}
{d\,Mv\over d\,t}=4\pi R_{\textsc{snr}}^2 P\,.
\ee
\end{enumerate}
Let \(t\) be the time since the pulsar's birth:
\be
t=t'-t_{b}'\,,
\ee
where the birth epoch is 
\be\label{eq:t_birth}
t_{b}'=-t_0'\left[1-\left({\omega\over\tilde\omega}\right)^{n-1}\right]\,.
\ee
Here, we express the pulsar's  current angular velocity and the velocity at the  
birth,  as fractions of the breakup velocity: \(\Omega_{p}=\omega\Omega^{(0)}_{\rm br}\) and \(\Omega_{b}=\tilde\omega\Omega^{(0)}_{\rm br}\), respectively.
The pulsar's time-dependent  spin-down luminosity is
\be\label{eq:L_birth}
L_{\rm sd}(t)={L_{p} \left(\tilde\omega/\omega\right)^{(n+1)}\over \left(1+t/ t_0\right)^{{n+1\over n-1}}}\,,
\ee
where the characteristic  slow-down time is
\be\label{eq:t0_birth}
t_0={2 \tau\over (n-1)}\left({\omega\over\tilde\omega}\right)^{n-1}\,.
\ee
Obviously, the present epoch corresponds to \(t_{p}=-t_{b}'\).
\subsubsection{Asymptotic Expansion}
For \(t\gg t_0\),  the pulsar injection ceases significantly, and the expansion of the SNR is dominated by the initial energy release.
At the initial stage of evolution of the PWN inside the composite SNR does not proceed adiabatically because of the energy injection by the pulsar \citep[see, e.g.,][]{1984ApJ...278..630R} and the severe radiative losses caused by the strong magnetic field \citep{1973ApL....13..103P}. However, at timescales exceeding the pulsar slow-down time, the intensity of the injection by the pulsar ceases. The extension of the nebula results in the reduction of the magnetic field and suppression of radiative losses. Thus, at the later stages of expansion, the PWN should evolve adiabatically. The approach suggested by \citet{1967pswh.book.....Z} can be generalized to describe the evolution of the PWN.
The energy of the system consisting of a PWN and the SNR shell is
\be
\begin{split}
E=&
{4\pi\over 3}\left({\rho_{\textsc{ism}} \dot{R}_{\textsc{snr}}^2\over2}R_{\textsc{snr}}^3+\right.\\
&\left.{3P\over2}\left(R_{\textsc{snr}}^3-R_{\textsc{pwn}}^3\right)+3PR_{\textsc{pwn}}^3\right)\,,
\end{split}
\ee
where the first and second terms describe the contribution from the shell,  and the third term characterizes the contribution from the PWN. At later stages,  the main contribution to the change of the kinetic energy of the shell is due to the increase of its mass, thus
\be\label{eq:pressure_simple}
P\propto \rho_{\textsc{ism}}\dot{R}_{\textsc{snr}}^2\,.
\ee
This allows us to express the energy of the system as
\be
E=2\pi P\left(R_{\textsc{pwn}}^3+kR_{\textsc{snr}}^3\right)\,,
\ee
where \(k\) is a dimensionless constant.
Since the expansion is adiabatic, \(PR_{\textsc{pwn}}^4=\rm const\). 
$E$ and $P$  are related as 
\be
2\pi PR_{\textsc{pwn}}^4=aE
\ee
or, equivalently,
\be\label{eq:R_snr_on_R_pwn}
\dot{R}_{\textsc{snr}}^2\propto a{E\over \rho_{\textsc{ism}}} R_{\textsc{pwn}}^{-4}\,.
\ee
Here, \(a\) is a constant of length dimension. The expression for the energy can be simplified as
\be
R_{\textsc{pwn}}^3\left(R_{\textsc{pwn}}-a\right)=kaR_{\textsc{snr}}^3\,.
\ee
Combining the above two equations, one obtains
\be
\dot{R}_{\textsc{pwn}}\propto a^{5/6}\sqrt{E\over\rho_{\textsc{ism}}}{\left(R_{\textsc{pwn}}-a\right)^{2/3}\over R_{\textsc{pwn}}\left(R_{\textsc{pwn}}-3a/4\right)}\,.
\ee
This equation has an analytic solution, which  in the asymptotic limit \(R_{\textsc{pwn}}\gg a\)
is reduced to 
\be \label{eq:rp_limit}
R_{\textsc{pwn}}\propto a^{1/4}\left({Et^2\over \rho_{\textsc{ism}}}\right)^{3/20}\,.
\ee
The radius of the PWN allows us to determine the shell radius and the pressure of the SNR
(see Eqs. \eqref{eq:R_snr_on_R_pwn} and \eqref{eq:pressure_simple}):
\be \label{eq:rs_limit}
R_{\textsc{snr}}\propto\left({Et^2\over \rho_{\textsc{ism}}}\right)^{1/5}
\ee
and
\be
P\propto \rho_{\textsc{ism}}\left({E\over \rho_{\textsc{ism}}t^3}\right)^{2/5}\,.
\ee
Finally, the ratio of the pulsar spin-down luminosity and the pressure determines 
the location of the pulsar wind TS:
\be \label{eq:rts_limit}
R_{\textsc{ts}}\propto\sqrt{L(t)\over P}\propto t^{\frac35-\frac{n+1}{2(n-1)}}\,.
\ee
Although the SNR expansion is similar to Sedov's solution (with a different
numerical coefficient),  the PWN expands slower with time: \(R_{\textsc{pwn}}\propto t^{0.3}\). To infer the details of the SNR expansion,  we solve the system of  dynamic equations numerically.
\subsubsection{Numerical Model}
The dynamic equations can be completed with the following 
phenomenological relations for the shell momentum:
\be \label{eq:shell_mom}
Mv=\frac{4\pi}3 R_{\textsc{snr}}^3 C_{m}\rho_{\textsc{ism}}{dR_{\textsc{snr}}\over dt}\,,
\ee
and the shell kinetic energy
\be \label{eq:shell_energy}
{Mv^2\over2}=\frac{4\pi}3R_{\textsc{snr}}^3C_{e}\rho_{\textsc{ism}}\left({d\,R_{\textsc{snr}}\over d\,t}\right)^2\,.
\ee
Adjusting the constants \(C_{m}\) and \(C_{e}\) provides a better agreement between the prediction of this simple model and the more accurate numerical treatment. In the case of a strong explosion in a medium  with adiabatic index \(5/3\), the blast-wave radius and pressure are
\be
R_{\textsc{snr}}=\left({\alpha Et^2\over \rho_{\textsc{ism}}}\right)^{1/5}
\ee
and
\be\label{eq:pressure_ratio}
P=\beta P_{\textsc{snr}}=\frac{3\beta}4\rho_{\textsc{ism}}\left({d\,R_{\textsc{snr}}\over d\,t}\right)^2\,,
\ee
where \(\alpha\simeq1.29\) and \(\beta=0.3\) are dimensionless coefficients.  The solution of the system correctly reproduces the blast-wave radius and the pressure if one adopts \(\dst C_m = \frac{3 \beta}{2} \approx 0.45\) and \(C_e= \frac{75}{16\pi\alpha}-\frac{9 \beta}{8} \approx 0.82\) (see Appendix~\ref{sec:appendix_SN} for details).
The system of  dynamic equations and phenomenological relations can be reduced to the following system:
\begin{equation} \label{eq:p_dot}
\prodi{P}{t} = \rho_{\textsc{ism}} \frac{5 x^2 A - 8 B}{(4+x^3) R_{\textsc{snr}}}~,
\end{equation}
\begin{equation}\label{eq:rpwn_dot}
\prodi{R_{\textsc{pwn}}}{t} = \rho_{\textsc{ism}} \frac{ (1-x^3) A + 2x B}{(4+x^3) P},
\end{equation}
\begin{equation}\label{eq:v_dot}
\prodi{V_{\textsc{snr}}}{t} = \frac{ 3\dfrac{P}{\rho_{\textsc{ism}}} - 3 C_m V_{\textsc{snr}}^2}{C_m R_{\textsc{snr}}},
\end{equation}
\begin{equation}\label{eq:rsnr_dot}
\prodi{R_{\textsc{snr}}}{t} = V_{\textsc{snr}}~.
\end{equation}
Here, the following notations are  used: 
\be
\dst x = \frac{R_{\textsc{pwn}}}{R_{\textsc{snr}}},\quad A = \frac{L}{4 \pi \rho_{\textsc{ism}} R_{\textsc{pwn}}^2}\,,
\ee
 and 
\be
B = V_{\textsc{snr}} \Big[ \left( \frac{2C_e}{C_m}+\frac{3}{2}
\right) \frac{P}{\rho_{\textsc{ism}}} - 3 C_e V_{\textsc{snr}}^2 \Big]\,.
\ee
Similarly to Sedov's solution, the inertia of the ISM affects the expansion of the SNR through the dimensionless parameter, 
\be
s=\sqrt{L_0 t_0^3\over \rho_{\textsc{ism}} R_0^5}\,,
\ee
where \(R_0\), \(t_0\), and \(L_0\) are the characteristic spatial, time, and energy scales (see Appendix~\ref{sec:appendix_dimless} for the dimensionless form of the equations). 
\subsubsection{Initial Conditions}
\begin{figure}
\resizebox{\columnwidth}{!}{
\includegraphics[]{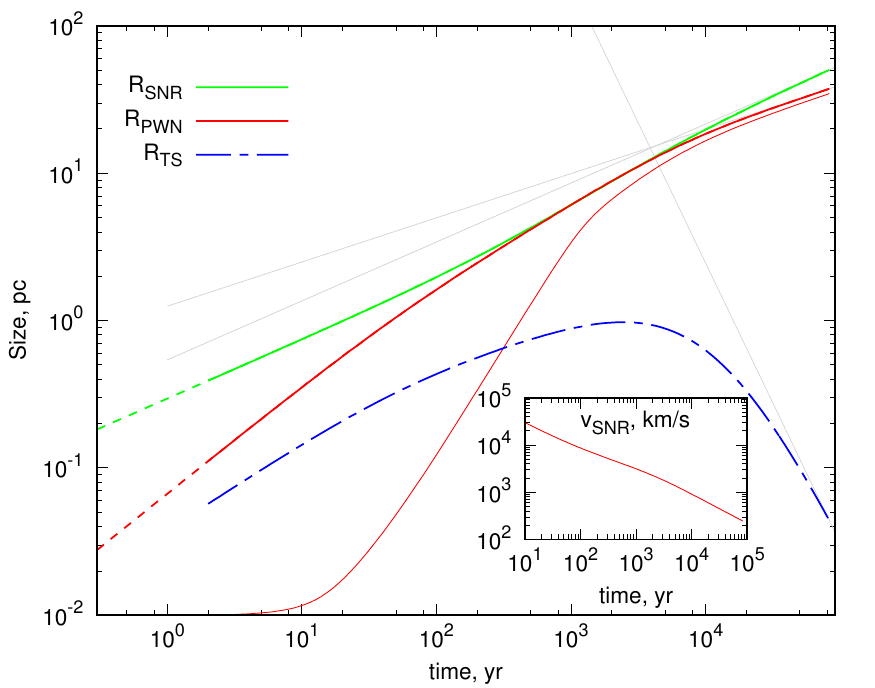}}
\caption{Numerical solution of the system of Equations~\eqref{eq:p_dot}~--~\eqref{eq:rsnr_dot} together with lines showing the asymptotic behavior as predicted by Equations~\eqref{eq:rs_limit}, \eqref{eq:rp_limit}, and \eqref{eq:rts_limit}. The pulsar angular velocity and the braking index  are assumed to be \({\tilde\omega}=0.5\) and \(n=1.5\), respectively. The calculation is performed for the ISM with a mean gas number density 
\({\bar n}_{\textsc{ism}}=1\rm\,cm^{-3}\). Thin line shows a solution of a system similar to Eqs.~\eqref{eq:p_dot}~--~\eqref{eq:rsnr_dot}, where the mass of the shell has an additional contribution of  \(30M_\odot\) ejecta. The inclusive plot shows the dependence of the SNR shock velocity.}
\label{fig:shock}
\end{figure}

Equations \eqref{eq:p_dot}~--~\eqref{eq:rsnr_dot} can be solved numerically, e.g., using the
Runge-Kutta method. To proceed with the integration, one needs to define the initial values for \(P\), \(R_{\textsc{pwn}}\), \(R_{\textsc{snr}}\), and \(V_{\textsc{snr}}\). The initial values related to the SNR shell should account for the energy transferred  at the SN explosion
to the ejecta
\be
\left.\frac{4\pi}3 \rho_{\textsc{ism}}R_{\textsc{snr}}^3V_{\textsc{snr}}^2\right|_{t=0}\simeq E_{\textsc{ej}}\,.
\ee
The initial radii of the PWN and SNR should satisfy the obvious condition  \(R_{\textsc{pwn}}<R_{\textsc{snr}}\ll 1\rm\,pc\). Also, the initial energy of the relativistic gas should be small 
compared to the characteristic energy injected by the pulsar:
\be
\left.\frac{4\pi}3 R_{\textsc{pwn}}^3P\right|_{t=0}\ll L_{b} t_{0} \,.
\ee
If these requirements are satisfied, the impact of the initial conditions on the solution of Eqs. \eqref{eq:p_dot}~--~\eqref{eq:rsnr_dot} is weak and vanishes once the energy injected by the pulsar significantly exceeds the initial energy.  The pulsar initial condition is determined by its birth angular velocity or, equivalently, by its birth epoch for the given braking index; see Eqs.~\eqref{eq:t_birth}~--~\eqref{eq:t0_birth}.
A typical solution of the system of Equations \eqref{eq:p_dot}~--~\eqref{eq:rsnr_dot} is shown in Fig.~\ref{fig:shock}. 

More detailed modeling should be performed with numerical simulations of (M)HD equations. This should allow an accurate
description of the pressure distribution through the system, which is described in our simplified model by
Eq.~\eqref{eq:pressure_ratio}. Detailed HD simulations should also allow us to account for the initial state of the ejecta
and resolving its fine structure. The hydrodynamic processes in the SN shell should provide a smaller impact on the size
of the system, as compared to the key processes accounted for in our simplified model.

The presence of massive ejecta should significantly slow down the expansion of the system at the earlier stages. This effect
can also be instigated with Eqs. \eqref{eq:p_dot}~--~\eqref{eq:rsnr_dot}, if the ejecta mass is added to the mass terms
in Eqs. \eqref{eq:shell_mom} and \eqref{eq:shell_energy}. As it is shown in Fig.~\ref{fig:shock}, for heavy ejecta with
\(M_\textsc{ej}=30M_\odot\), its influence is significant for the first \(10^3\rm\,yr\) and becomes negligibly small
after a few thousands of years.

In the frameworks of the considered dynamic model, the initial state of ejecta has a tiny impact on the final size the
nebula. However, the small size of the nebula at the initial stages enhances the rate of the radiative losses
significantly. If the radiative losses are dominant, the expansion of the nebula proceeds in a nonadiabatic regime, as
it was assumed in Eq.~\eqref{eq:eq1}, and the nebula may not be able to push the SN shell as efficient as
predicted by Eqs. \eqref{eq:p_dot}~--~\eqref{eq:rsnr_dot}.

The rate of radiative losses is determined by the composition of the pulsar wind and its magnetization. Conventionally,
one assumes that electrons/positrons are the predominant ingredients of pulsar winds. There is a significant uncertainty
in the value of the electron mean Lorentz factor, \(\G\). It can range from a few hundred (similar to the pulsar wind Lorentz
factor close to the light cylinder) to a value of \(\sim10^6\) anticipated in the Crab Nebula.  The magnetic field tends
to the equitation with the gas pressure; however, initially, the magnetic field can be weak, with the magnetic pressure
being just a few percent of the gas pressure.

\begin{figure}
\resizebox{\columnwidth}{!}{
\includegraphics[]{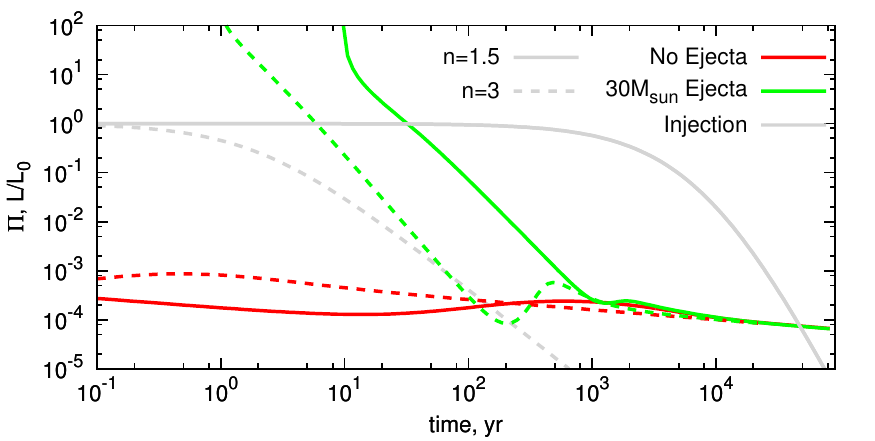}}
\caption{ Hydrodynamic parameter \(\Pi\) that determines the ratio of the synchrotron and adiabatic losses, \(\k\),  through Eq.~\eqref{eq:losses_ratio}.  Calculations are performed for two values of the braking index: \(n=1.5\) (solid lines) and \(n=3\) (dashed lines). A heavy ejecta (\(M_\textsc{ej}=30M_\odot\)) significantly enhances the rate of radiative cooling at the early stage of expansion. Gray lines show the normalized rate of injection by the pulsar.}
\label{fig:cooling}
\end{figure}

To compare the rate of synchrotron and adiabatic cooling, we assume that the magnetic pressure is equal to a fraction of
the gas pressure: \(P_{\textsc{b}}=(B^2/8\pi)=\k_{\textsc{b}}P\). The ratio of the losses is
  \be
  \label{eq:losses_ratio} \k = {\dot{\gamma}_{\rm syn} \over \dot{\gamma}_{\rm ad}}=
-{\frac43c\sigma_\textsc{t}P_{\textsc{b}}\G^2\over \frac13\G{d\ln n\over
    dt}}
=\Pi \k_\textsc{b}\G\,.
\ee
The function \(\Pi\) depends only on the hydrodynamic properties of the nebula, thus can be evaluated with the model
described by Eqs. \eqref{eq:p_dot}~--~\eqref{eq:rsnr_dot}.  In Fig.~\ref{fig:cooling} the parameter \(\Pi\) is shown as
a function of time since the pulsar birth. For large values, \(\Pi > 10^{-2}\), the nebula expansion most likely
proceeds in a nonadiabatic regime, thus the obtained solution has limited applicability. A massive ejecta causes the
nonadiabatic regime at the initial stage. Depending on the braking index, the transition to the adiabatic regime should
occur between tens or hundreds of years. In the case of large values of the braking index, \(n\sim3\), the pulsar
slow-down occurs on a timescale shorter than the duration of the nonadiabatic phase, thus the solution obtained with
Eqs. \eqref{eq:p_dot}~--~\eqref{eq:rsnr_dot} may overestimate the size of the PWN. In contrast, for the small values of
the braking index, \(n\sim 1.5\), the pulsar slow-down takes significantly longer than the duration of the nonadiabatic
phase, thus the obtained solution should provide a meaningful estimate on the size of the nebula.

{ Radiative cooling may affect the dynamics of the nonrelativistic shell.  Matter emissivity depends on its temperature, the abundances of chemical elements, and the density. Several simple approximations for the radiative losses have been obtained by a number of authors \citep[see, e.g.,][and references therein]{2008MNRAS.388..357K}. The  model described by Eqs.~(\ref{eq:shell_mom}~-~\ref{eq:rsnr_dot}) does not allow an accurate evaluation of the radiative cooling since such an estimate requires a detailed information about the density and temperature distributions in the nonrelativistic shell. Thus, to address the radiative cooling we compute a fraction of energy radiated in the case of Sedov's explosion. Accounting for the density and temperature distributions, the radiated energy fraction is}
\be
\begin{split}
  \frac{\Delta E}{E}<\quad & 8\times10^{-3} \left({n_{\textsc{ism}}\over 1 \rm\,cm^{-3}}\right)^{1.64}\times\\
  &\left({E\over 10^{51}\rm\,erg}\right)^{-0.64} \left({t\over 10\rm\,kyr}\right)^{2.92}\,.
\end{split}
\ee
{  The  emitted energy, \(\Delta E\), corresponds to radiative losses in the entire shell integrated from the explosion moment to \(t\) (see Appendix~\ref{sec:appendix_cooling} for detail). From this estimate one can conclude that for }\(n_\textsc{ism}\sim 1\rm\,cm^{-3}\) and \(E\gg10^{51}\rm\,erg\) {  the shock should remain nonradiative for the entire evolution of the system, \(t\leq10^5\rm\,yr\).}
\section{Implications for  \pwn}\label{sec:application}
\subsection{Size of the nebula}
\begin{figure}
\resizebox{\columnwidth}{!}{
\includegraphics[]{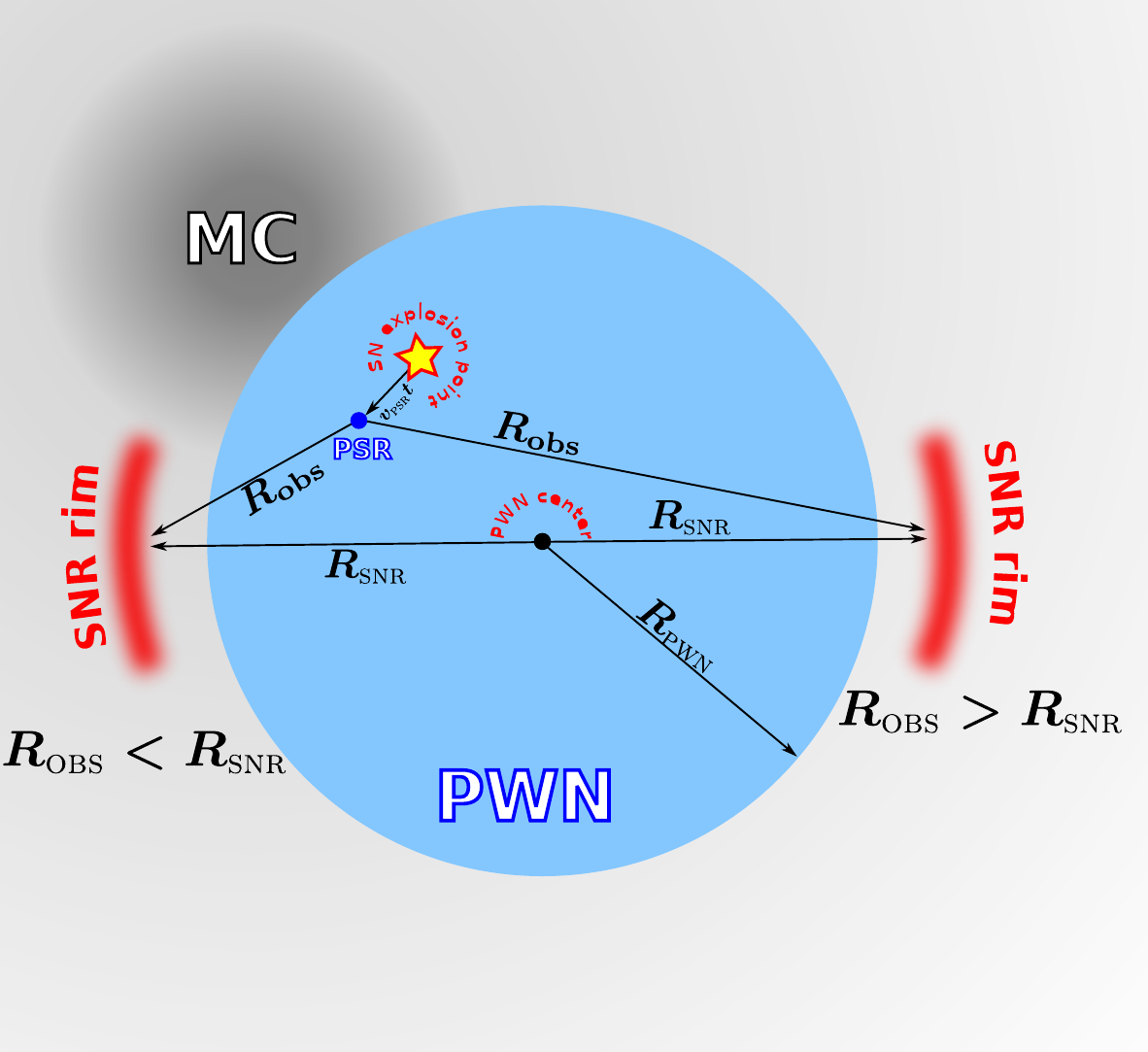}} 
\caption{ Arrangement and characteristic distances in the 1D model. In the frameworks of 1D model, all the relevant distances, \(R_\textsc{pwn}\) and \(R_\textsc{snr}\),  should be measured from the geometrical center of the nebula. A nearby molecular cloud (indicated as ``MC'') can render an anisotropic expansion of the system. }
\label{fig:j1825}
\end{figure}

For the distance to the source of \pwn  of  4~kpc \citep[based on dispersion measure using the models for
Galactic electrons by][]{1993ApJ...411..674T,2002astro.ph..7156C}, the 
measured  angular size of the latter corresponds to the 70~pc linear size of the gamma-ray production region. Although the  gamma-ray image  deviates from the spherically symmetric shape; below for simplicity  we will assume that the radius of the source is 35~pc.
The detection of a H$\alpha$ rim at a distance of \(R_{\textsc{obs}}=120\rm\,pc\) from the pulsar PSR J1826-1334
has been  interpreted as a signature of the progenitor SNR  \citep[see][]{2008MNRAS.390.1037S,2016MNRAS.458.2813V}.
{ As illustrated in Fig.~\ref{fig:j1825}, in the case of an ISM density gradient, the measured radius does not coincide with the radius, }\(R_\textsc{snr}\).{ If the pulsar is still located inside the SNR then the SNR radius cannot be significantly smaller than the observed distance: \(R_\textsc{snr}>R_\textsc{obs}/2\). Thus, with a factor of two uncertainty one can estimate the radius based on Sedov's solution:}
\be
\begin{split}
R_{\textsc{snr}}&\simeq\left({1.3E_{\textsc{ej}}t_{p}^2\over \rho_{\textsc{ism}}}\right)^{1/5}\\
&\sim12\left({E_{\textsc{ej}}\over 10^{51}\erg}\right)^{1/5}\left({t_{p}\over10\rm\,kyr}\right)^{2/5}\left({n_{\textsc{ism}}\over 1\rm \,cm^{-3}}\right)^{-1/5}\rm \,pc\,,\\
\end{split}
\ee
where \(\rho_\textsc{ims} = n_\textsc{ism}\, m_p\) are the density, the hydrogen number density, and the proton mass, respectively.
{ Thus, the SNR shell can be located at the distance of \(R_\textsc{obs}\sim120\)~pc,  provided that the following condition is fulfilled: }
{\be
    \left({t_p\over 10\rm\,kyr}\right)^{2}\left({E_{\textsc{ej}}\over10^{51}\erg}\right)\left({{\bar n}_{\textsc{ism}}\over 1\rm\,cm^{-3}}\right)^{-1}>3\times10^3\,, 
\ee}
where the numerical value corresponds to \(R_\textsc{snr}= R_\textsc{obs}/2\simeq60\rm\,pc\).
The ejecta energy at SN type II is constrained by \(E_{\textsc{ej}}\simeq3\times10^{51}\erg\), while the age of the source is limited by \(t_p< 100\,\rm kyr\) (for \(n\simeq1.5\)).  Thus, Sedov's solution is consistent with the observed size of the SNR when {\({\bar n}_{\textsc{ism}}<0.1\rm\,cm^{-3}\)}. Note that  \cite{2009ASSL..357..451D} derived somewhat smaller upper limit  for ${\bar n}_{\rm ism}$ assuming the braking index $n=2$.  This upper limit seems to be below the typical density of the interstellar gas in the Galactic Plane. Moreover,  \citet{2016MNRAS.458.2813V} reported the presence of several molecular clouds in the small angular proximity to the nebula with an average number density  $n_{\rm c}\sim10^3 \, \rm cm^{-3}$ and characteristic size 
 \(r_{\rm c}\sim10\rm\,pc\).  It is likely that one of these clouds is responsible for the anisotropic expansion of the nebula, thus it should be located close to the nebula. The gas density in  molecular clouds is expected to follow the King's profile:
\be
\rho={\rho_{\rm c}\over (1+r/r_{\rm c})^2}\,.
\ee
Thus, the density to the south of the pulsar can hardly be as low as it is required by Sedov's solution.  Instead, it  should be comparable to the upper limit of \(n_\textsc{ism}\sim4\rm\,cm^{-3}\) obtained by \citet{2016MNRAS.458.2813V}.

To avoid this problem, we suggest an alternative  scenario in which  the energy injection by the pulsar  dominates over the kinetic energy of the ejecta.  This scenario has an advantage since it offers significantly larger available energy. 

The energy for the source expansion is transferred through the PWN, which evolves adiabatically, and the energy, consumed for the expansion, is determined by the volumes of the PWN and SNR. Thus, their sizes should be measured neither from the SN explosion point nor from the present pulsar location. Instead one should use, as a starting point,  the geometrical center of the current structure.

Below, we adopt the following requirements to be met by the 1D model:
\(R_{\textsc{ts}}\sim0.03\rm\,pc\),  \(R_{\rm pwn}\sim35\rm\,pc\), 
and \(R_{\textsc{snr}}>R_{\textsc{obs}}/2=60\rm\,pc\).  As it follows from  Figure~\ref{fig:j1825}, the 3D structure of the system implies a large uncertainty for the relation between \(R_{\textsc{snr}}\)  and \(R_{\textsc{obs}}\).  This can be consistently resolved only with realistic 3D numerical simulations, which is beyond the scope of this paper.

\begin{figure}
\resizebox{\columnwidth}{!}{
\includegraphics[]{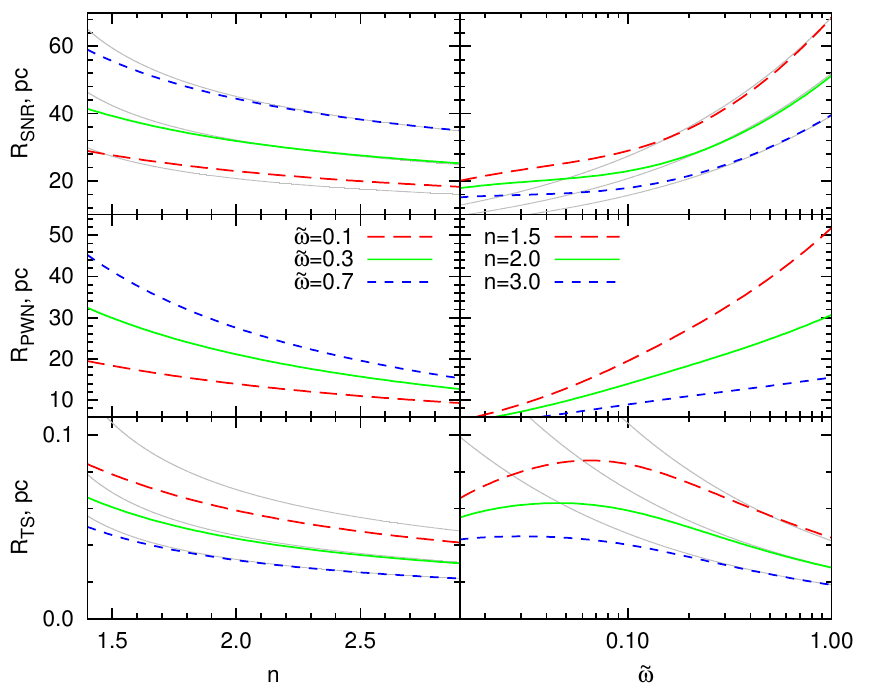}}
\caption{Present-day radii of the SNR (top panels), PWN (middle panels), and pulsar wind TS (bottom panels) as functions of the pulsar braking index \(n\) (left panels) and pulsar birth angular velocity (right panels). In the left panels, the  calculations are performed for three values of the pulsar angular velocity at its birth: \(\tilde\omega=0.1\), \(0.3\), and \(0.7\). These numbers correspond to 1\%, 10\%, and 50\% of the pulsar gravitational energy. In the right panels, the calculations are performed  for three values of the braking index \(n=1.5\), \(2\), and \(3\). The adopted number density of the surrounding gas is  \({{\bar n}_{\textsc{ism}}}= 1\rm\,cm^{-3}\). The thin gray lines show analytic approximations given by Eqs. \eqref{eq:snr} and \eqref{eq:ts}.}
\label{fig:high_density}
\end{figure}
\begin{figure}
\resizebox{\columnwidth}{!}{
\includegraphics[]{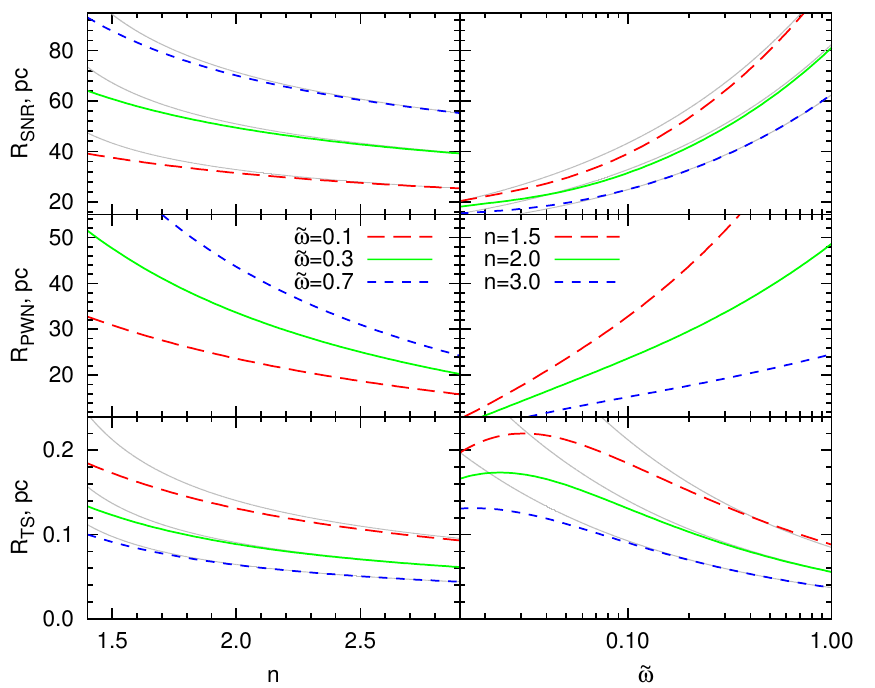}}
\caption{Same as in Fig.~\ref{fig:low_density} but for \({{\bar n}_{\textsc{ism}}}=0.1 \rm\,cm^{-3}\).}
\label{fig:low_density}
\end{figure}
For the  present-day spin-down luminosity and the characteristic age of the pulsar, the size of the the composite SNR is determined by three parameters: the braking index \(n\), the initial rotation velocity \(\tilde{\omega}\), and the density of the surrounding medium \({\bar\rho}_{\textsc{ism}}\).   The braking index and the initial rotation velocity determine the age of the system through Eq.~\eqref{eq:t_birth}, the birth spin-down luminosity through Eq.~\eqref{eq:L_birth}, and the initial slow-down time through Eq.~\eqref{eq:t0_birth}.

The results of the calculation of the present-day size are shown in Figs.~\ref{fig:high_density} and \ref{fig:low_density}.  These calculations allow determination of the numerical coefficients for the asymptotic expressions given by Eqs.~(\ref{eq:rs_limit}) and (\ref{eq:rts_limit}):
\be
\label{eq:snr}
R_{\textsc{snr}}\simeq 52 \left({\tilde{\omega}^2\over (n-1)^2\rho_0}\right)^{1/5}\rm\,pc\,,
\ee
\be
\label{eq:ts}
R_{\textsc{ts}}\simeq 0.028 \left({(n-1)^6\tilde{\omega}^4\rho_0^{3} }\right)^{-1/10}\rm\,pc\,,
\ee
where \(\rho_0={{\bar \rho}_{\textsc{ism}}}/(m_p \rm\,cm^{-3})\). For \({\tilde\omega}>0.1\) and \(n> 1.5\),  these approximations are rather accurate;  it can be checked from the comparison  of the results of the numerical calculations shown in Figs.~\ref{fig:high_density} and \ref{fig:low_density}.  For small  values of \(\tilde\omega\leq0.1\),  the pulsar contribution to the overall energy budget becomes comparable to the assumed initial  energy of the SNR shell, which was ignored when deriving these expressions. Also, it is assumed that the bulk of the pulsar rotation energy is promptly  released at the initial phase. This assumption violates for small values of the braking index \(n\). Finally, we note that because of the presence of parameter \(a\) in Eq.~\eqref{eq:rp_limit}, the size of the PWN cannot be described  by an approximate formula  similar to Eqs.~\eqref{eq:snr} and \eqref{eq:ts}.
A comparison of Eq.~\eqref{eq:ts} with the TS radius implies that the density of the medium should be rather high, \({\bar n}_{\textsc{ism}}\sim 1\rm\,cm^{-3}\). For this high medium density, the SNR and PWN can be sufficiently extended if \({\tilde\omega}>0.5\) and \(n\leq 2\) as shown in Fig.~\ref{fig:high_density}.
\section{Summary}\label{sec:conclusion}
The total energy released at an SN type II explosion, \(10^{53}\erg\), exceeds, by two orders of magnitude, the energy
transferred to the ejecta, \(E_{\textsc{ej}}\sim10^{51}\erg\). The bulk of energy escapes with the neutrino
emission. Typically, the pulsar rotation energy is small compared to the ejecta energy. However, it cannot be excluded
that at some circumstances the pulsar can receive a larger fraction of the explosion energy. When it becomes comparable
to the kinetic energy of the ejecta, the sizes of both the SNR and the PWN can significantly deviate from Sedov's
solution. To illustrate the implication of this scenario, we considered a specific case of \pwn, a very extended and
bright in VHE gamma-rays PWN \citep{2006A&A...460..365A,2008ApJ...675..683P,2017AIPC.1792d0035M,2017arXiv170803126M,2018A&A...612A...2H}.
The X-ray and gamma-ray observations of \pwn show that the shape of the nebula deviates from the spherically symmetric
geometry. Thus, the 1D model described in Sec.~\ref{sec:pwn} has limited capability for accurate quantitative
predictions.  However, in the framework of the suggested scenario, the hydrodynamic processes evolve differently from
the conventional situations, and the 1D model should provide a correct estimate for the energy required for the PWN/SNR
inflation.  In the framework of the scenario, the pressure in the PWN is the main driving force of the SNR
expansion. Independently on symmetry of the system and dimension of the used model, the PWN is nearly isobaric, thus the
radius of the pulsar wind TS provides an observational constraint for the pressure in the
nebula. Therefore, we adopt the following requirement: the present-day pressure in the nebula should be consistent with
the pulsar wind TS located at \(R_{\textsc{ts}}\sim0.03\rm\,pc\) from the pulsar
\citep{2011ApJ...742...62V}.  The 1D model allows us to obtain the radii of the SNR, PWN, and pulsar wind TS as functions of three key model parameters: density of the ISM, pulsar braking index, and the initial angular
velocity of the pulsar. In the framework of this model, it is possible to reproduce the three key properties of the
system: \(R_{\textsc{snr}}\geq60\rm\,pc\), \(R_{\textsc{pwn}}\simeq35\rm\,pc\), and
\(R_{\textsc{ts}}\simeq0.03\rm\,pc\), provided that (i) the pulsar obtains, at its birth, a significant fraction of the
explosion energy (\({\tilde\omega}\geq0.5\)), and (ii) the braking index is small, \(n\leq2\). { The conventional Sedov-like solution implies a low density of the ISM,}
\({\bar n}_{\textsc{ism}}<0.1\rm\,cm^{-3}\). { In contrast,  the model proposed in this paper requires a rather dense environment,} \({\bar n}_{\textsc{ism}}\sim 1\rm \,cm^{-3}\).  { This agrees with the presence of dense
  molecular clouds reported in the vicinity of the source} \citep{2016MNRAS.458.2813V}. What concerns the requirement
for the pulsar braking index, \(n\leq2\), is that it is small compared to the value measured for the Crab pulsar
\citep{1993MNRAS.265.1003L},  but it matches both the old \citep[e.g., Vela;
\(n=1.4\pm0.2\),][]{1996Natur.381..497L} and young \citep[e.g., J1833\(-\)1034;
\(n=1.8569\pm0.001\),][]{2012MNRAS.424.2213R} pulsars.

{ Equations~(\ref{eq:snr})~and~(\ref{eq:ts}) allow us to estimate the model parameters based on the requirement to reproduce the observed sizes of the SNR and the position of the TS. For example, if the true size of the SNR approaches \(120\rm\,pc\),  then the scenario suggested in this paper requires} \((n-1)\rho_0^{1/2}\simeq0.43\) (for \(R_\textsc{ts}\simeq0.03\rm\,pc\)).  { Formally, this relation can be easily fulfilled. However, one should also account for the required birth angular velocity:} \(\tilde{\omega}\simeq3.5\). { Such a high value implies (see Eq.~(\ref{eq:e_rot})) that the pulsar physical parameters significantly deviates from the conventional values.}

The most demanding requirement for the suggested scenario is the
initial rotation energy of the pulsar. Our simulations require fast angular velocity at the birth,
\({\tilde\omega}\geq0.5\), which implies a very small initial rotation period, \(P_{\rm b}\leq1\rm\,ms\). This translates
to a lower limit on the birth spin-down luminosity 
\be
 \label{eq:sd_range} 
 L_{\rm b} > 3\times10^{41}\ergs\,, \ee for the braking index and the initial angular velocity: \(n =1.5\) and
 \({\tilde\omega}=0.5\), respectively.  The requirement for the very fast initial rotation may seem an extreme
 assumption. However, as \citet{2005ApJ...626..350H} have shown, the differential rotation in a massive progenitor
 should result in the formation of a rapidly rotating pulsar. For example, numerical simulations predict the birth of a pulsar
 with \(P_{\rm b}=3\rm\,ms\) for a \(35M_\odot\) progenitor star.  Thus, the option of an initially very fast rotating
 pulsar {\it a priori} cannot be excluded. It should be considered along with the conventional scenario that provides an
 explanation of the very large size of \pwn assuming an extremely low density environment surrounding the source.

 \acknowledgments 
The work of F.~Aharonian, S.V.~Bogovalov, and G.V.~Ustuygova was supported by the Russian Science
Foundation, project No. 16-12-10443. Authors are grateful to the anonymous referee for useful comments that helped to improve the manuscript.
\appendix
\section{Numerical Model for the SN Explosion}\label{sec:appendix_SN}
If the injection of energy by the PWN is sufficiently  small to have an impact  
on  the dynamics of the SNR, the dynamic equations consist of 
the requirement  of {\it energy conservation}:
\begin{equation} \label{5.1}
\prodi{}{t} \left( \frac{M v^2}{2} + \frac{3PV_{\textsc{snr}}}{2} \right) = 0~.\,;
\end{equation}
and the {\it  Second Newton law}:
\begin{equation}
\prodi{M v}{t} = 4 \pi R_{\textsc{snr}}^2 P~. \label{5.2}
\end{equation}
Here, it is adopted that the shell mass is concentrated close to the blast-wave.
Eqs.~(\ref{5.1}) and (\ref{5.2}) can be completed with empirical relations for the shell's momentum and kinetic energy:
\[
Mv = C_m \rho_{\textsc{ism}} \frac{4 \pi R_{\textsc{snr}}^3}{3} \prodi{R_{\textsc{snr}}}{t},~~~~\frac{M v^2}{2} = C_e \rho_{\textsc{ism}} \frac{4 \pi
R_{\textsc{snr}}^3}{3} \left( \prodi{R_{\textsc{snr}}}{t} \right)^2~.
\]
The solution of the hydrodynamics equation allows us to find  numerically the  coefficients for Sedov's self-similar solution. Namely, the blast-wave radius, $R_{\textsc{snr}}$, and pressure at explosion point, $P_c$, are 
\[
R_{\textsc{snr}} = \left( \frac{\alpha E t^2}{\rho_{\textsc{ism}}} \right)^{1/5},~~~~P = \beta P_c = \frac{3\beta}{4} \rho_{\textsc{ism}}
\dot R_{\textsc{snr}}^2~,
\]
where \(E\) and \(P\) are the explosion energy and pressure behind the blast-wave. For polytropic gas with \(\gamma=5/3\), the numerical coefficients are  \(\alpha\approx 1.29,~\beta \approx 0.3\).
Eq.~(\ref{5.1}) can be integrated, yielding
\begin{equation}
\frac{M v^2}{2} + \frac{3PV}{2} = \frac{4\pi}{3} \rho_{\textsc{ism}} C_e R_{\textsc{snr}}^3 \dot R_{\textsc{snr}}^2 + \frac{4\pi}{3} R_{\textsc{snr}}^3
\frac{3P}{2} = \frac{4\pi}{3} R_{\textsc{snr}}^3  \left( \rho_{\textsc{ism}} C_e \dot R_{\textsc{snr}}^2 +  \frac{3P}{2} \right) = E~.
\label{5.3}
\end{equation}
In turn, Eq.~(\ref{5.3}) is fulfilled if
\begin{equation}
\dot R_{\textsc{snr}} = A R_{\textsc{snr}}^{-3/2}~, \label{5.4}
\end{equation}
\begin{equation}
P = B R_{\textsc{snr}}^{-3}~. \label{5.5}
\end{equation}
Note that
\begin{equation}
\frac{4\pi}{3} \left( \rho_{\textsc{ism}} C_e A^2 +  \frac{3B}{2} \right) = E~.\label{5.6}
\end{equation}
Solving Eq.~(\ref{5.4}) one obtains \(\dst R_{\textsc{snr}} = \left( \frac{5At}{2} \right)^{2/5} = \left(
\frac{\alpha E t^2}{\rho_{\textsc{ism}}} \right)^{1/5},~~~~\frac{25 A^2}{4} = \frac{\alpha E}{\rho_{\textsc{ism}}}\), which yields in
\(\dst A^2 = \frac{4 \alpha E}{25 \rho_{\textsc{ism}}}\).
Changing variables in Eq.~(\ref{5.2}) as $R_{\textsc{snr}},~~~~d R_{\textsc{snr}} = D dt,~~~~D =\dot R_{\textsc{snr}}$, one obtains
\begin{equation}
\frac{4\pi}{3} \rho_{\textsc{ism}} C_m \prodi{R_{\textsc{snr}}^3 D}{R_{\textsc{snr}}} = 4 \pi \frac{R_{\textsc{snr}}^2 P}{D}~.\label{5.7}
\end{equation}
Substituting in Eq.~(\ref{5.7}) one derives that $\dst \frac{\rho_{\textsc{ism}} C_m A^2}{2}  = B$. Thus, the following chain of equations should be satisfied 
\[
\frac{P}{\dot R_{\textsc{snr}}^2} = \frac{3 \beta \rho_{\textsc{ism}}}{4} = \frac{B}{A^2} = \frac{\rho_{\textsc{ism}} C_m}{2}~.
\]
This implies $\dst C_m = \frac{3 \beta}{2}$.
Eq.~(\ref{5.6}) can be simplified as
\begin{equation}
\frac{3 E}{4\pi} = \rho_{\textsc{ism}} C_e A^2 +  \frac{3 C_m \rho_{\textsc{ism}} A^2}{4} = \rho_{\textsc{ism}} \left( C_e  +  \frac{3
C_m}{4} \right) \frac{4 \alpha E}{25 \rho_{\textsc{ism}}} =  \left( C_e  +  \frac{3 C_m}{4} \right) \frac{4
\alpha E}{25}~,\label{5.8}
\end{equation}
and, consequently,
\begin{equation}
\frac{3}{4\pi} = \left( C_e  +  \frac{3 C_m}{4} \right) \frac{4 \alpha}{25}~.\label{5.9}
\end{equation}
This allows us to obtain the coefficient as
\begin{equation}
C_e = \frac{75}{16 \pi \alpha} - \frac{3 C_m}{4} = \frac{75}{16 \pi \alpha} - \frac{9 \beta}{8}~.
\end{equation}
This means that the dynamic model can correctly reproduce the size of the SNR and the pressure of the gas at the explosion
point if \(\dst C_m = \frac{3 \beta}{2} \approx 0.45,~~C_e = \frac{75}{16\pi\alpha}-\frac{9\beta}{8} \approx
0.82\). The parameters \(\alpha\approx 1.29\) and \(\beta\approx 0.3\) are the numerical coefficients for Sedov's
solution in polytropic gas with \(\gamma=5/3\). The obtained values for \(C_e\) and \(C_m\) are also used in the dynamic model for the composite  ``PWN+SNR'' model.
\section{Dimensionless Form of the Equations}\label{sec:appendix_dimless}
The  dimensionless equations that determine the expansion of the SNR and PWN are
\begin{equation}
\prodi{p}{\tau} = \frac{5 x^2 a - 8 (b_1+b_2/s)}{(4+x^3) r_{\textsc{snr}}}~,
\end{equation}
\begin{equation}
\prodi{r_{\textsc{pwn}}}{\tau} = \frac{ (1-x^3) a + 2x (b_1+b_2/s)}{(4+x^3) p},
\end{equation}
\begin{equation}
\prodi{v_{\textsc{snr}}}{t} = \frac{ \dfrac{3ps}{C_m } - 3 v_{\textsc{snr}}^2}{r_{\textsc{snr}}},
\end{equation}
\begin{equation}
\prodi{r_{\textsc{snr}}}{\tau} = v_{\textsc{snr}}~.
\end{equation}
Here, the characteristic time, space and energy scales---\(t_0\), \(R_0\), and \(L_0\)---define the dimensionless variables  \(t=t_0\tau\), \(R_{{\textsc{snr}}/pwn}=R_0r_{{\textsc{snr}}/pwn}\), \(V_{\textsc{snr}}=v_{\textsc{snr}}(R_0/t_0)\), \(P={L_0t_0\over R_0^3}p\),  \(\dst x = \frac{r_{\textsc{pwn}}}{r_{\textsc{snr}}}\), \(a = \frac{L}{4 \pi L_0 r_{\textsc{pwn}}^2}\), \(b_1 = v_{\textsc{snr}} p \left( \frac{2C_e}{C_m}+\frac{3}{2}
\right)\), and \(b_2 = - C_e v_{\textsc{snr}}^3\). Similarly to Sedov's solution, a dimensionless parameter
\be
s={L_0t_0^3\over \rho_{\textsc{ism}}R_0^5}
\ee
accounts for the inertia of the ISM. 
\section{Radiation Cooling for Sedov's Solution}\label{sec:appendix_cooling}
Thermal radiation may affect properties of the nonrelativistic shock.  The nonrelativistic shell should be quite unhomogeneous and the physical conditions in the shell undergo substantial changes during the SNR evolution. Since the thermal emission coefficient depends nonlinearly on the plasma density and temperature, one should utilize a precise spatial--time description of the nonrelativistic gas to address the thermal cooling accurately.  The model suggested in this paper shares much in common with the classic Sedov solution, especially in regard to the evolution of the nonrelativistic gas. Thus, to evaluate the importance of the thermal cooling, we calculate a fraction of radiated energy for a self-similar Sedov's explosion. 

We adopt the adiabatic index of ideal nonrelativistic gas to be \(\gamma=5/3\). Temperature and density in the self-similar solution depend on the spatial coordinate as 
\be \label{eq:tn_profiles}
T = T_s F(\xi),~~~n = n_s G(\xi)\,,
\ee
where \(\xi = {r/ R_\textsc{snr}(t)}\) is the dimensionless radius. Functions in Eq.~(\ref{eq:tn_profiles}) selected to satisfy the following conditions: \(F(1)=G(1)=1\). The shock-front values of density and temperatures are
\be
n_s = 4 n_\textsc{ism},~~~~T_s = \frac{p_s}{2 k_B n_s} = \frac{3 m_p}{200 k_B} \Big( \frac{R_\textsc{snr}}{t} \Big)^2\,,
\ee
where the shock-front pressure is
\be
p_s = \frac{3 \rho_0}{4} \Big( \frac{2R_\textsc{snr}}{5t} \Big)^2 = \frac{3 m_p n_\textsc{ism}}{25} \Big(
\frac{R_\textsc{snr}}{t} \Big)^2\,.
\ee

Thermal radiation losses are determined by the cooling function \(\Lambda\) as 
\be
\begin{split}
\prodi{E}{t} &= \int\limits_0^{R_\textsc{snr}} n_\textsc{h} n_e \Lambda(T) 4\pi r^2dr = 4\pi A \int\limits_0^{R_\textsc{snr}} n^2 T^{-a} r^2dr = 4\pi A
R_\textsc{snr}^3 n_s^2 T_s^{-a} \int\limits_0^1 G(\xi)^2 F(\xi)^{-a} \xi^2d\xi\\
&= 4\pi A R_\textsc{snr}^3 16 n_\textsc{ism}^2 \Big( \frac{3 m_p}{200 k_B} \Big)^{-a} \Big( \frac{R_\textsc{snr}}{t} \Big)^{-2a} \int\limits_0^1
G^2 F^{-a} \xi^2 d\xi\,,
\end{split}
\ee
where one assumed hydrogen gas, i.e., the electron density, \(n_e\), is equal to the proton density, \(n_\textsc{h}\). The TS radius is 
\be
R_\textsc{snr} = \Big( \frac{E t^2}{m_p n_\textsc{ism}} \Big)^{1/5}\,,
\ee
and the cooling function was approximated as \(\Lambda=AT^{-a}\).
This results in the following thermal radiation losses:
\be
\begin{split}
\prodi{E}{t} &= 64\pi A \Big( \frac{E t^2}{m_p n_\textsc{ism}} \Big)^{3/5} n_\textsc{ism}^2 \Big( \frac{3 m_p}{200 k_B}
\Big)^{-a} \Big( \frac{E}{m_p n_\textsc{ism}} \Big)^{-2a/5} t^{6a/5} \int\limits_0^1 G^2 F^{-a} \xi^2 d\xi\\
&= C A \Big( \frac{E}{m_p} \Big)^{(3-2a)/5} n_\textsc{ism}^{1.4+2a/5} \Big( \frac{3 m_p}{200 k_B} \Big)^{-a}
t^{6(1+a)/5}\,,
\end{split}
\ee
where \(C = 64\pi \int\limits_0^1 G^2 F^{-a} \xi^2 d\xi\) is a constant.
According to \citet{1979ApJS...41..555H}, one can adopt \(a=0.6\), thus the thermal cooling is 
\be
\prodi{E}{t} = C A \Big( \frac{E}{m_p} \Big)^{0.36} n_\textsc{ism}^{1.64} \Big( \frac{3 m_p}{200 k_B}
\Big)^{-0.6} t^{1.92}.
\ee
The total thermal losses are obtained by integration over time in the limits from $0$ to $t$:
\be
\begin{split}
\frac{\Delta E}{E} &= \frac{C A}{2.92 E} \Big( \frac{E}{m_p} \Big)^{0.36} n_\textsc{ism}^{1.64} \Big( \frac{3
  m_p}{200 k_B} \Big)^{-0.6} t^{2.92} \\
&\approx 7.7 \times 10^{14} C \left({A\over \rm \erg\,cm^3\,s^{-1}\,K^{0.6}}\right) \left({E\over 10^{51} \erg}\right)^{-0.64}
\left({n_\textsc{ism}\over 1\rm\,cm^{-3}}\right)^{1.64} \left({t\over10\,\rm kyr}\right)^{2.92}.
\end{split}
\ee
Constant \(C\) contains a large factor \(64\pi\) and a small integral term. For the \(\xi<1\) function \(G\) rapidly decreases from the shock-front value \(G(1)=1\). In contrast, function \(F\) rapidly increases from the shock-front value \(F(1)=1\). Mass conservation implies that 
\be
\int\limits_0^1 G \xi^2 d\xi = 1/12,
\ee
thus \(C < 64 \pi /12 < 16.8\), and \(C \approx 10\) seems to be a reasonable estimate. Accounting for \(A = 6.2 \times 10^{-19}\rm\,\erg\,cm^3\,s^{-1}K^{0.6}\) from \citet{1979ApJS...41..555H}; a similar value is obtained for the cooling function approximation by \citet{2007ApJS..168..213G}, one obtains 
\be
\begin{split}
\frac{\Delta E}{E} &< 7.7 \times 10^{14} \times 16.8 \times 6.2 \times 10^{-19} \left({E\over 10^{51} \erg}\right)^{-0.64}
\left({n_\textsc{ism}\over 1\rm\,cm^{-3}}\right)^{1.64} \left({t\over10\,\rm kyr}\right)^{2.92} \\
&< 8 \times 10^{-3} \left({E\over 10^{51} \erg}\right)^{-0.64}
\left({n_\textsc{ism}\over 1\rm\,cm^{-3}}\right)^{1.64} \left({t\over10\,\rm kyr}\right)^{2.92}\,.
\end{split}
\ee.

%\bibliographystyle{apj} 
%\bibliography{bibliography}

\end{document}